\documentstyle[sprocl,epsf]{article}

\def\Journal#1#2#3#4{{#1} {\bf #2}, #3 (#4)}

\def\NPB{{\em Nucl. Phys.} B}
\def\NPA{{\em Nucl. Phys.} A}
\def\PLB{{\em Phys. Lett.}  B}
\def\PRL{\em Phys. Rev. Lett.}

\def\PRC{{\em Phys. Rev.} C}

\def\beq{\begin{equation}}
\def\eeq{\end{equation}}
\def\beqa{\begin{eqnarray}}
\def\eeqa{\end{eqnarray}}
\newcommand{\boldtau}{\mbox{\boldmath $\tau$}}
\newcommand{\boldS}{\mbox{\boldmath $S$}}

\begin{document}

\title{\hfill TRI-PP-99-04\\[0.4cm]
EFFECTIVE THEORY FOR NEUTRON-DEUTERON SCATTERING AND THE TRITON}

\author{H.-W. HAMMER}

\address{TRIUMF, 4004 Wesbrook Mall,
Vancouver, B.C.\\ Canada V6T 2A3\\E-mail: hammer@triumf.ca}


\maketitle\abstracts{We apply the effective field theory approach
to the three-nucleon system. In particular, we consider neutron-deuteron
scattering and the triton. Precise predictions for $S=3/2$
scattering are obtained in a straightforward way.
In the $S=1/2$ channel, however, a unique 
nonperturbative renormalization takes place which requires the 
introduction of a three-body force at leading order.
We also show that invariance under the renormalization group
explains some universal features of the three-nucleon system.}

\section{Introduction}
Effective field theories (EFT) are a powerful concept
designed to explore a separation of scales in physical 
systems.\cite{gospel} For example if the momenta $k$ of two
particles are much smaller than the inverse range of 
their interaction $1/R$, observables can be expanded in powers of $kR$.
Following the early work of Weinberg,\cite{weinberg}
EFT's have become quite popular in nuclear physics.\cite{nn98,monster} 
Their application, however, is complicated
by the presence of shallow (quasi) bound states, which create 
a large scattering length $a\gg R$. In this finely tuned 
case, the perturbative expansion in $kR$ already
breaks down at rather small momenta $k\sim1/a$. In order to 
describe bound states with typical momenta $k\sim1/a$,
the range of the EFT has to be extended.
A certain class of diagrams has to be resummed which generates a 
new expansion in $kR$ where powers of $ka$ are kept to all orders.
For the two-nucleon system a power counting that incorporates
this resummation has been found recently.\cite{1stooge,3musketeers,gegelia}
Pions are included and treated perturbatively in this scheme. It has 
successfully been applied to $NN$ scattering and deuteron
physics.\cite{martin}

The three-nucleon system is a natural testground for the understanding
of the nuclear forces that has been reached in the two-nucleon system.
However, the extension of these ideas is not straightforward, as
the three-nucleon system shows some remarkable universal features.
It has been found that different
models of the two-nucleon interaction that are fitted to the same
low-energy two-nucleon ($NN$) data predict different but correlated
values of the triton binding energy $B_3$ and the
$S$-wave nucleon-deuteron ($Nd$) scattering length $a_3^{(1/2)}$ 
in the spin $S=1/2$ channel;
all models fall on a line in the $B_3 \times a_3^{(1/2)}$ plane,
the Phillips line.\cite{phillips}
Other universal features of three-body systems are
the existence of a logarithmic spectrum of bound states
that accumulates at zero energy as the 
two-particle scattering length $a_2$ increases
(the Efimov effect \cite{efimov}), and the collapse of 
the deepest bound state 
when the range of the two-body interaction $r_2$ goes to zero 
(the Thomas effect \cite{thomas}).

We will show that the renormalization of the three-nucleon
system is nonperturbative and requires a three-body force at leading
order. In addition to the description of experimental data from a few
low-energy parameters, EFT allows to understand the above
mentioned universal features in a unified way.\cite{paulo,bosons,baryons} 

\section{Neutron-Deuteron System}
In order to avoid the difficulties due to the long range Coulomb force,
we concentrate here on the neutron-deuteron ($nd$) system. We also 
restrict ourselves to scattering below the deuteron breakup threshold
where $S$-waves are dominant. There is only one scale at low energy,
$\sqrt{MB_2}\approx 40$ MeV, where $B_2$ is the binding energy of the 
deuteron and $M$ the nucleon mass. Since $\sqrt{MB_2}$ is small compared 
to the pion mass, the pions can be integrated out and only the nucleon 
degrees of freedom remain. The leading order 
of this pionless theory is equivalent to the leading order
in the KSW counting scheme \cite{3musketeers} where pions are included
perturbatively. As a consequence, the extension of our results to higher 
energies is well defined. Recently, a number of two nucleon observables 
have been studied in the pionless theory as well.\cite{chen}

There are two $S$-wave channels for neutron-deuteron scattering,
corresponding to total spin $S=3/2$ and $S=1/2$. For scattering 
in the $S=3/2$ channel all spins are aligned and the two-nucleon
interactions are only in the $^3 S_1$ partial wave. The interaction
is repulsive and the Pauli principle forbids the three nucleons to be at 
the same point in space. As a consequence, this channel is insensitive 
to short distance physics and very precise predictions are obtained 
in a straightforward way. There is also no three-body bound state in 
this channel. The $S=1/2$ channel is more complicated.
The two-nucleon interaction can take place either in the $^3 S_1$
or in the $^1 S_0$ partial waves. This leads to an attractive interaction 
which sustains a three-body bound state, the triton. The $S=1/2$ channel
also shows a strong sensitivity to short distance physics as the Pauli
principle does not apply. As a consequence, 
it displays the Thomas \cite{thomas} and Efimov \cite{efimov} effects.
The generic features of this channel are very similar to the system of 
three spinless bosons. There is a strong cutoff dependence even though
all Feynman diagrams are finite. As we will show, the renormalization 
requires a leading order three-body force counterterm.\cite{bosons,baryons}
Since the details of the renormalization in the three-body system are
discussed in Bedaque's talk,\cite{paulo} we will mainly focus on
the application of these ideas to the $nd$ system.

\begin{figure}[htb]
\begin{center}
\epsfxsize=10cm
\centerline{\epsffile{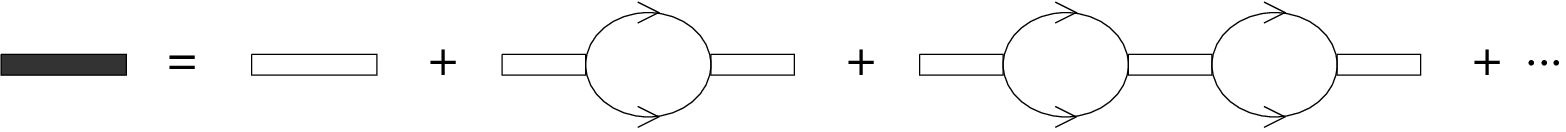}}
\end{center}
\caption{Dressing of the bare dibaryon propagator.}
\label{fig:dress}
\end{figure}

Let us start from the assumption that the three-body force is
of natural size. The lowest order effective Lagrangian is then given by 
\beqa
\label{lagN}
{\cal L}&=&N^\dagger \left(i\partial_0 +\frac{\vec{\nabla}^2}{2M}\right)N
-C_0^t\left(N^T \tau_2 \vec{\sigma} \sigma_2 N\right)^\dagger \cdot\left(
N^T \tau_2 \vec{\sigma}\sigma_2 N\right) \\
& &-C_0^s\left(N^T \sigma_2 \boldtau \tau_2 N\right)^\dagger \cdot \left(
N^T \sigma_2 \boldtau \tau_2 N\right) +\ldots\,,\nonumber
\eeqa
where the dots represent higher order terms supressed by derivatives
and more nucleon fields.
$\vec{\sigma}$ $(\boldtau)$ are Pauli matrices operating in
spin (isospin) space, respectively.
The contact terms proportional to $C_0^t$ ($C_0^s$) correspond to 
two-nucleon interactions in the $^3 S_1$ ($^1 S_0$) $NN$ channels.
Their renormalized values are related to the 
corresponding two-body scattering lengths $a_2^t$ and $a_2^s$ by 
$C_0^{s,t}=\pi a_2^{s,t}/(2M)$.
Since no derivative interactions are included, this 
Lagrangian generates only two-nucleon interactions of zero range.
For practical purposes, it is convenient to rewrite this theory 
by introducing \lq\lq dibaryon'' fields
with the quantum numbers of two nucleons.\cite{david}
In our case, we need two dibaryon fields: (i) a field $\vec{T}$ with
spin (isospin) 1 (0) representing two nucleons interacting in the $^3 S_1$
channel (the deuteron) and (ii) a field $\boldS$ with
spin (isospin) 0 (1) representing two nucleons interacting in the $^1 S_0$
channel.
% which only possesses a virtual bound state.
Using a Gaussian path integration, it is straightforward to show
that the Lagrangian from Eq. (\ref{lagN}) is equivalent to
\beqa
\label{lagd}
{\cal L}&=&N^\dagger \left(i\partial_0 +\frac{\vec{\nabla}^2}{2M}\right)N
+ \Delta_T \vec{T}^\dagger \cdot\vec{T} +\Delta_S \boldS^\dagger \cdot\boldS\\ 
& &-\frac{g_T}{2}\left( \vec{T}^\dagger \cdot N^T \tau_2 \vec{\sigma} 
\sigma_2 N +h.c.\right) -\frac{g_S}{2}\left(\boldS^\dagger \cdot N^T 
\sigma_2 \boldtau \tau_2 N +h.c.\right) +\ldots\,. \nonumber
\eeqa
At first it may look like the Lagrangian, Eq. (\ref{lagd}), contains more
parameters than the original one, Eq. (\ref{lagN}). However, 
the scales $\Delta_T$ and $\Delta_S$ are arbitrary and included in 
Eq. (\ref{lagd}) only to give the dibaryon fields the usual mass dimension 
of a heavy field. They can easily be removed by rescaling the dibaryon fields.
All observables depend only on the ratios $g^2_{T,S}/\Delta_{T,S}$.
Since the theory is nonrelativistic, all particles propagate forward in time,
the nucleon tadpoles vanish, and the propagator for the nucleon fields is
\beq
\label{nucprop}
iS(p)=\frac{i}{p_0-p^2/2M +i\epsilon}\,.
\eeq
The dibaryon propagators are more complicated because of the coupling
to two-nucleon states. The bare dibaryon propagator is simply a constant, 
$i/\Delta_{S,T}$, but the full propagator gets dressed by nucleon loops
to all orders as illustrated in Fig. \ref{fig:dress}.
The nucleon loop integral has a linear UV divergence which can be absorbed 
in $g^2_{T,S}/\Delta_{T,S}$, a finite piece determined by the unitarity 
cut, and subleading terms that have already been omitted in Eq. (\ref{lagd}). 
Summing the resulting geometric series leads to
\beq
\label{Dprop}
i D_{S,T}(p) =  \frac {-i}{- \Delta_{S,T}
             + \frac{M g_{S,T}^{2}}{2\pi} 
               \sqrt{-M p^0+\frac{\vec{p}^{\,2}}{4}-i\epsilon} +i\epsilon}\,,
\eeq
where $g_{T,S}$ and $\Delta_{T,S}$ now denote the renormalized parameters.
The $NN$ scattering amplitude in the respective channel is obtained by
attaching external nucleon lines to the dressed propagator.
In the center of mass frame, the $S$-wave amplitude in the 
$^3 S_1$, $^1 S_0$ channels for energy $E=k^2/M$ is
\beq
\label{Tamp}
^{3,1}T(k)=\frac{4\pi}{M}\left( -\frac{2\pi\Delta_{S,T}}{M  g_{S,T}^{2}}
-ik \right)^{-1}\,,
\eeq
and the renormalized parameters $g_{T,S}$ and $\Delta_{T,S}$ can
be determined from 
\beq
\label{renpar}
a_2^{s,t}=\frac{M  g_{S,T}^{2}}{2\pi\Delta_{S,T}}\,.
\eeq

\section{$S=3/2$ $nd$-Scattering}
\begin{figure}[htb]
\begin{center}
\epsfxsize=10cm
\centerline{\epsffile{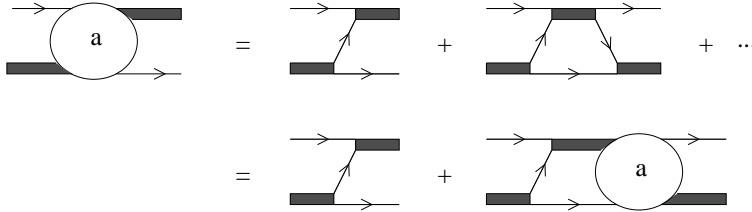}}
\end{center}
\caption{The first few diagrams contributing to $S=3/2$ $nd$-scattering 
(first line) and the integral equation which sums the diagrams to all 
orders (second line).}
\label{fig:32eq}
\end{figure}

\begin{figure}[htb]
\begin{center}
\epsfxsize=10cm
\epsffile{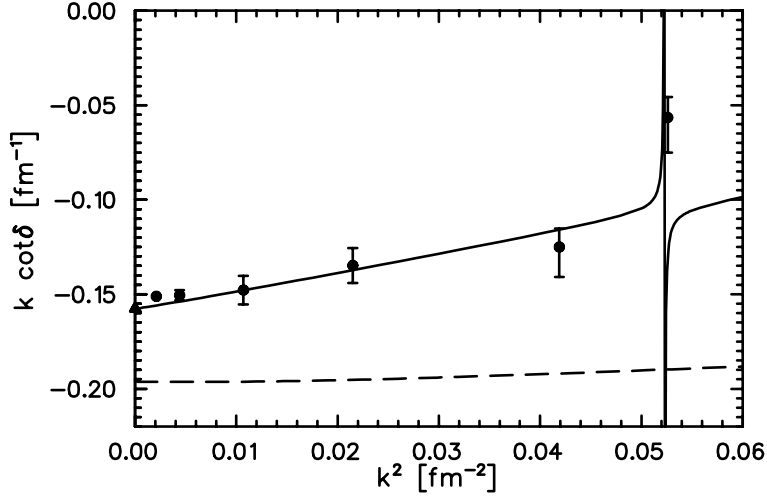}
\end{center}
\caption{$k\cot\delta$ for $S=3/2$ $nd$-scattering to order 
$(r_2/a_2)^0$ (dashed line) and $(r_2/a_2)^2$ (solid line). 
The dots \protect\cite{vOers} and the triangle \protect\cite{dilg} indicate 
the experimental values.}
\label{fig:res32}
\end{figure}

For $S=3/2$ scattering only the dibaryon field $\vec{T}$ contributes.
The first few diagrams that contribute are shown in the first line
of Fig. \ref{fig:32eq}.
Since the leading piece of all diagrams is of the order $\sim M g_T^2/Q^2$, 
they have to be resummed for typical momenta $Q\sim 1/a_2^t$.
This is conveniently achieved by solving the integral 
equation given by the second line in Fig. \ref{fig:32eq}.
Performing the integration over the time component of the four-momentum
flowing in the loop and projecting onto the $S$-waves, we obtain the 
scattering amplitude $a(k,p)$,\cite{skorny}
\beqa
\label{inteq32}
& &\frac{3}{4}\left(1/a^t_2+\sqrt{3 p^2/4-ME}\right)^{-1} a(p,k)\\
& &=-K(p,k)-\frac{2}{\pi}\int_0^\infty \frac{q^2\;dq}{
q^2-k^2-i\epsilon} K(p,q)\, a(q,k)\,,\nonumber
\eeqa
where
\beq
\label{2bkern}
K(p,q)=\frac{1}{2pq}\ln\left(\frac{q^2+pq-p^2-ME}{q^2-pq-p^2-ME} \right)\,
\eeq
and $ME=3 k^2/4-1/(a^t_2)^2$ is the total energy. 
%$a_2^t$ is the two-body scattering length in the $^3 S_1$ channel.
$k\,(p)$ denote the incoming (outgoing) momenta in the center
of mass frame. The amplitude $a(p,k)$ is normalized such that the 
on-shell value $a(k,k)=(k\cot\delta-ik)^{-1}$ where $\delta$ is the  
elastic scattering phase shift.
The solution of Eq. (\ref{inteq32}) can be obtained numerically
and is very insensitive to the high momentum 
modes in the integral equation. If, for example, the integral equation
is cut off at some finite momentum $\Lambda$, the low energy behavior of 
the solution remains unchanged. The EFT in this channel is very 
predictive. The solution of Eq. (\ref{inteq32}) is shown by the dashed 
line in Fig. \ref{fig:res32}. 
When the corrections of the effective range in the 
two-nucleon interaction are taken into account up to ${\cal O}
((r_2/a_2)^2)$, one obtains the 
solid curve \cite{3stooges} which agrees nicely with the phase shift analysis 
of van Oers and Seagrave (dots).\cite{vOers}
For the scattering length the agreement is even better. The calculation
to ${\cal O}((r_2/a_2)^2)$ gives \cite{2stooges} $a_3^{(3/2)}=(6.33 
\pm 0.10)$ fm to be compared with the experimental value \cite{dilg}
$a_3^{(3/2)}=(6.35 \pm 0.02)$ fm. Furthermore, the EFT expansion in 
powers of $r_2/a_2$ converges well, as the contributions
from ${\cal O}((r_2/a_2)^0)$ to ${\cal O}((r_2/a_2)^2)$
to the scattering length are $a_3^{(3/2)}=(5.09 + 0.91 + 0.33)$ fm, 
in order. The extension beyond the deuteron breakup threshold and
the inclusion of pions in KSW counting has recently been carried out by 
Bedaque and Grie{\ss}hammer.\cite{begrie}

\section{$S=1/2$ $nd$-Scattering and the Triton}
\begin{figure}[htb]
\begin{center}
\epsfxsize=10cm
\centerline{\epsffile{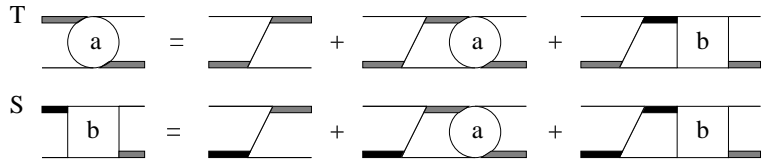}}
\end{center}
\caption{Coupled integral equations for $S=1/2$ $nd$-scattering.
$\vec{T}\,(\boldS)$ dibaryon is indicated by grey shaded (black) thick
line.}
\label{fig:eq12}
\end{figure}

In the $S=1/2$ channel, the situation is more complicated. The 
two-nucleon interactions can now take place both in the $^3S_1$ and $^1S_0$ 
partial waves. As shown in Fig. \ref{fig:eq12}, there are
two coupled amplitudes, $a$ and $b$.
The amplitude $a$ which has both an incoming and outgoing dibaryon
field $\vec{T}$ gives the phase shifts for $S=1/2$ $nd$-scattering. 
However, $a$ is coupled to the amplitude $b$ which has an 
incoming dibaryon $\boldS$ and an outgoing dibaryon $\vec{T}$. 
Although only $a$ corresponds to $S=1/2$ $nd$-scattering,
both amplitudes have the quantum numbers of the triton.
From the Lagrangian, Eq. (\ref{lagd}),
one obtains the coupled integral equations for the two amplitudes.
After the integration over the time component of the loop-momentum
and the projection onto the $S$-waves has been carried out, we have
\begin{eqnarray}
\label{aeq}
& &\frac{3}{2}\left(1/a_2^t+\sqrt{3 p^2 /4 -ME}\right)^{-1}a(p,k)\\
& &=K(p,k)+\frac{2}{\pi}\int_0^\Lambda\frac{q^2 \;dq}{q^2-k^2-i\epsilon}
K(p,q) [a(q,k)+3 b(q,k)] \nonumber\\
\label{beq}
& & 2\,\frac{\sqrt{3 p^2 /4 -ME}-1/a_2^s}{p^2-k^2}\;b(p,k)\\
& &=3 K(p,k)+\frac{2}{\pi}\int_0^\Lambda\frac{q^2 \;dq}{q^2-k^2-i\epsilon}
K(p,q) [3 a(q,k)+b(q,k)]\,.\nonumber
\end{eqnarray}
As before,
$k$ ($p$) denote the incoming (outgoing) momenta in the center of mass 
frame and $M E = 3k^2/4 - (1/a_2^t)^2$ is the total energy.
The kernel $K(p,q)$ is given in Eq. (\ref{2bkern}).
%\beq
%K(p,q)=\frac{1}{2pq}\ln\left(\frac{q^2+pq-p^2-ME}{q^2-pq-p^2-ME}\right)\,,
%\eeq
$a_2^t$ $(a_2^s)$ are the scattering lengths
in the $^3S_1$ $(^1S_0)$ $NN$ channels, respectively.
The amplitude $a(p,k)$ is normalized such that $a(k,k)=
(k\cot\delta -ik)^{-1}$ with $\delta$ the elastic scattering phase shift
in the $S=1/2$ channel. Furthermore, we
have introduced a momentum cutoff $\Lambda$ in the integral equations.
Eqs. (\ref{aeq}, \ref{beq}) have previously been derived 
using different methods.\cite{skorny} In the 
limit $\Lambda\to \infty$ these equations do not have a unique solution
because the phase of the asymptotic solution is undetermined.\cite{danilov}
For a finite $\Lambda$ this phase is fixed and the solution is
unique. However, the equations with a cutoff have 
the same disease as in the boson case: a strong cutoff dependence that does 
not appear in any order perturbation theory.\cite{paulo,bosons,baryons} The 
amplitude $a(p,k=\mbox{const.})$ shows a strongly oscillating behavior. 
Varying the cutoff $\Lambda$ slightly changes the asymptotic phase 
by a number of ${\cal O}(1)$ and results in large changes of the 
amplitude at the on-shell point $a(p=k)$. This cutoff dependence 
is not created by divergent Feynman diagrams. It is a nonperturbative 
effect and appears although all individual diagrams are UV finite.

\begin{figure}[htb]
\begin{center}
\epsfxsize=10cm
\epsffile{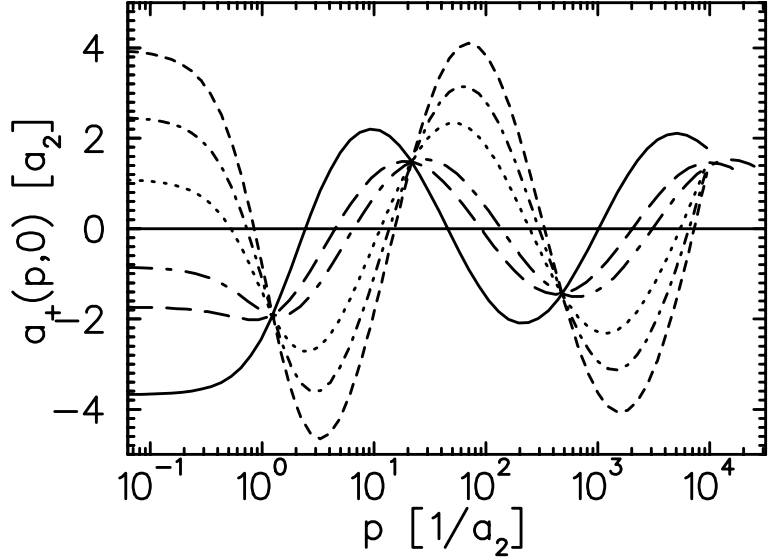}
\end{center}
\caption{Cutoff dependence of $a_+(p,k=0)$. 
Solid, dashed, and dash-dotted curves are
for $H=0$ and $\Lambda=1.0,\,2.0,\,3.0\times 10^4 a_2^{-1}$, respectively.
Dotted, short-dash-dotted, and short-dashed curves show the effect of the 
three-body force for $\Lambda=10^4 a_2^{-1}$ and $H=-6.0,\,-2.5,\,-1.8$, 
respectively.}
\label{fig:aplus}
\end{figure}

In order to control this strong $\Lambda$ dependence, 
it is useful to note that the integral equations (\ref{aeq},
\ref{beq}) are $SU(4)$ symmetric in the ultraviolet (UV). 
Therefore it is sufficient to consider the $SU(4)$ limit ($a_2^t=a_2^s=a_2$),
since the cutoff dependence is a problem rooted in the 
UV behavior of the amplitudes. Furthermore, we note
that the equations for $a_+=[a+b]$ and $a_-=[a-b]$ decouple in the $SU(4)$ 
limit. The equations for $a_+$ and $a_-$ are given by
\beqa
\label{apeq}
& &\frac{3}{4}\left(1/a_2+\sqrt{3 p^2/4-ME}\right)^{-1}a_+(p,k) \\
& &= 2K(p,k)+\frac{2H(\Lambda)}{\Lambda^2}
+\frac{2}{\pi}\int_0^\Lambda \frac{q^2\;dq}{q^2-k^2-i\epsilon}
   \left\{ 2K(p,q) + \frac{2H(\Lambda)}{\Lambda^2}\right\}a_+(q,k)\nonumber\\
\label{ameq}
& &\frac{3}{4}\left(1/a_2+\sqrt{3 p^2/4-ME}\right)^{-1}a_-(p,k)\\
& &=-K(p,k)\vphantom{\frac{1}{2}}
-\frac{2}{\pi}\int_0^\Lambda \frac{q^2\;dq}{q^2-k^2-i\epsilon}
   K(p,q)\, a_-(q,k)\nonumber\,,
\eeqa
where we have introduced a contact three-body force
$H(\Lambda)$ which runs with the cutoff $\Lambda$ into the equation for 
$a_+$. Let us disregard the three-body force for a moment. The equation
for $a_+$ is exactly the same equation as in the case of spinless
bosons while the equation for $a_-$ is the same equation as in the 
$S=3/2$ channel. The $S=3/2$ equation is well behaved and its
solution is very insensitive to the cutoff as discussed in the previous
section. Consequently, the observed cutoff dependence stems solely from the 
equation for $a_+$. As an example, the cutoff dependence of $a_+(p,k=0)$ 
is shown by the solid, dashed, and dash-dotted curves
in Fig. \ref{fig:aplus} for three different cutoffs, $\Lambda=1.0,\,
2.0,\,3.0\times 10^4 a_2^{-1}$. 
But we already know the solution to this problem: a three-body force 
counterterm that runs with the cutoff $\Lambda$.\cite{bosons} 
This is exactly the three-body force
we have introduced into Eq. (\ref{apeq}). The dotted, short-dash-dotted, 
and short-dashed curves in Fig. \ref{fig:aplus} show the effect of the 
three-body force $H(\Lambda)$ on $a_+(p,k=0)$ for $\Lambda=10^4 a_2^{-1}$ 
and $H=-6.0,\,-2.5,\,-1.8$. It is clearly seen that the variation of 
$H$ for a constant $\Lambda$ has the same effect on the amplitude as
varying the cutoff. Consequently, we can compensate the 
changes in the asymptotic phase when $\Lambda$ is varied by adjusting 
the three-body force term appropriately. (A more detailed discussion of 
the renormalization procedure can be found in Bedaque's talk.\cite{paulo})

\begin{figure}[htb]
\begin{center}
\epsfxsize=10cm
\centerline{\epsffile{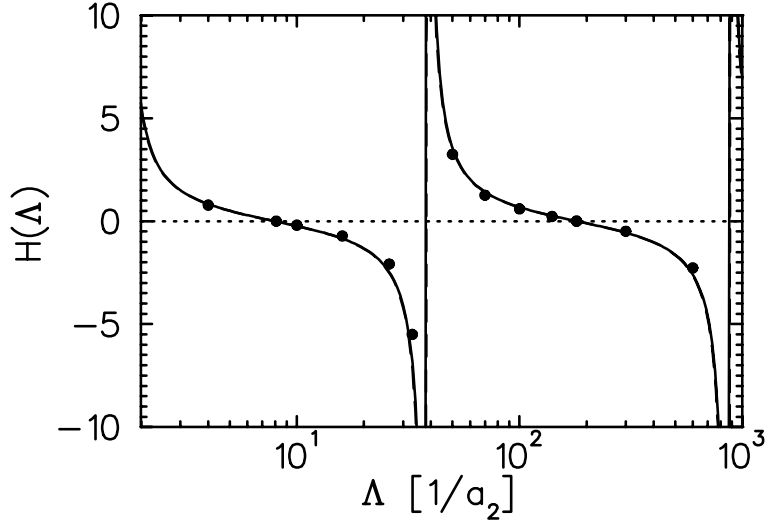}}
\end{center}
\caption{Running of $H(\Lambda)$ for $\Lambda_* = 0.9 \mbox{ fm}^{-1}$:
(a) from Eq. (\ref{runH}) (solid line), (b) from numerical solution of
Eq. (\ref{apeq}) (dots).}
\label{fig:betaf12}
\end{figure}

We can obtain an approximate expression for the running of $H(\Lambda)$ 
from invariance under the renormalization group. Requiring that the 
equation for $a_+$ does not change its form when the high momentum modes 
are integrated out, we find
\beq
\label{runH}
H(\Lambda)=-    \frac{\sin(s_0\ln({\Lambda}/{\Lambda_\star})-
                   {\rm arctg}(1/s_0))}
                 {\sin(s_0 \ln({\Lambda}/{\Lambda_\star})+
                   {\rm arctg}(1/s_0))}\,, 
\eeq
where $s_0\approx 1.0064$.\cite{paulo,bosons,baryons}
$H(\Lambda)$ contains one new dimensionful parameter, $\Lambda_*$,
which must be determined from experiment.
The running of the three-body force $H(\Lambda)$ according to
Eq. (\ref{runH}) is shown by the solid line in Fig. \ref{fig:betaf12}.
The dots are obtained by adjusting $H(\Lambda)$ such that the low
energy solution of Eq. (\ref{apeq}) remains unchanged when $\Lambda$
is varied. The observed agreement provides a numerical justification
for our proceeding. The three-body force
is periodic with $H(\Lambda_n)=H(\Lambda)$ for
$\Lambda_n=\Lambda\exp(n\pi/s_0)\approx \Lambda (22.7)^n$.
Since it enters only in the equation for $a_+=a+b$, 
the three-body force is also $SU(4)$ symmetric. 

Formally, the three-body force term in Eq. (\ref{apeq}) is obtained by adding 
\beqa
\label{3bod}
{\cal L}_3 = -\frac{2MH(\Lambda)}{\Lambda^2}\bigg(& & g_T^2 N^\dagger 
(\vec{T}\cdot\vec{\sigma})^\dagger (\vec{T}\cdot\vec{\sigma}) N \\
& & +\frac{1}{3} g_T g_S \left[ N^\dagger (\vec{T}\cdot\vec{\sigma})^\dagger
(\boldS\cdot\boldtau) N + h.c. \right] \nonumber \\
& & + g_S^2 N^\dagger (\boldS\cdot\boldtau)^\dagger (\boldS\cdot
\boldtau) N \bigg)\,,\nonumber
\eeqa
to the Lagrangian, Eq. (\ref{lagd}).
Eq. (\ref{3bod}) represents a contact three-body force written in
terms of dibaryon and nucleon fields. Via a Gaussian
path integration it is equivalent to a true three-nucleon force,
\beqa
\label{3bodN}
{\cal L}_3 = &-&\frac{2MH(\Lambda)}{\Lambda^2}\bigg(\,
\frac{g_T^4}{4\Delta_T^2}\,(N^T \tau_2 \sigma_k \sigma_2 N)^\dagger
(N^\dagger \sigma_k \sigma_l N)(N^T \tau_2 \sigma_l \sigma_2 N)\\
& &+\frac{1}{3}\frac{g_T^2}{2\Delta_T}\frac{g_S^2}{2\Delta_S}
\left[ (N^T \tau_2 \sigma_k \sigma_2 N)^\dagger
(N^\dagger \sigma_k \tau_l N)(N^T \sigma_2 \tau_l \tau_2 N) +h.c. \right]
\nonumber\\
& &+\frac{g_S^4}{4\Delta_S^2}\,(N^T \sigma_2 \tau_k \tau_2 N)^\dagger
(N^\dagger \tau_k \tau_l N)(N^T \sigma_2 \tau_l \tau_2 N)\,\bigg)\,.
\nonumber
\eeqa
By performing a Fierz rearrangement, it can then be shown that the three 
terms in Eq. (\ref{3bodN}) are equivalent. As a consequence, there is
only one three-body force which is also $SU(4)$ symmetric.
Naive power counting would suggest that the three-nucleon force 
scales with $1/(M m_\pi^4)$. The three-body force from Eqs. 
(\ref{3bod}, \ref{3bodN}), however, is enhanced by the renormalization 
group flow by two powers of $a_2$. Using Eq. (\ref{renpar}), it is found to  
scale as $a_2^2/(M m_\pi^2)$ which makes it leading order. 

Recently Mehen, Stewart, and Wise \cite{caltech}
found an approximate $SU(4)$ symmetry in the two-nucleon 
sector for $1/a_2 \ll p \ll m_\pi$ and also noticed that the only
$S$-wave four-nucleon force that can be written down
is $SU(4)$ symmetric.
Furthermore, there are no contact interactions with more than four 
nucleons without derivatives because of the Pauli principle.
It is also reasonable to assume that the low-energy dynamics of nuclei
is dominated by $S$-wave interactions. Therefore our 
$SU(4)$ symmetric three-body force together with the findings of
Mehen et al.\cite{caltech} gives an explanation 
for the approximate Wigner $SU(4)$ symmetry in nuclei.\cite{wigner}

Now we are in the position to solve the full equations for the 
broken $SU(4)$ case. Introducing  the three-body force from above
into Eqs. (\ref{aeq}, \ref{beq}), we obtain
\begin{eqnarray}
\label{aeqf}
& &\frac{3}{2}\left(1/a_2^t+\sqrt{3 p^2 /4 -ME}\right)^{-1}a(p,k)=
K(p,k)+\frac{2H(\Lambda)}{\Lambda^2}\\
& &+\frac{2}{\pi}\int_0^\Lambda\frac{q^2 \;dq}{q^2-k^2-i\epsilon}
\left[K(p,q) [a(q,k)+3 b(q,k)] +\frac{2H(\Lambda)}{\Lambda^2}[a(q,k)+b(q,k)]
\right]\nonumber\\
\label{beqf}
& & 2\,\frac{\sqrt{3 p^2 /4 -ME}-1/a_2^s}{p^2-k^2}\;b(p,k)=
3 K(p,k)+\frac{2H(\Lambda)}{\Lambda^2}\\
& &+\frac{2}{\pi}\int_0^\Lambda\frac{q^2 \;dq}{q^2-k^2-i\epsilon}
\left[K(p,q) [3 a(q,k)+b(q,k)] +\frac{2H(\Lambda)}{\Lambda^2}[a(q,k)+
b(q,k)]\right]\,.
\nonumber
\end{eqnarray}
We need one three-body datum to fix the three-nucleon force parameter 
$\Lambda_*$. We choose the experimental value for the $S=1/2$ $nd$-scattering 
length,\cite{dilg} $a_3^{(1/2)}=(0.65 \pm 0.04)\mbox{ fm}$
and find $\Lambda_* = 0.9 \mbox{ fm}^{-1}$. (For the special cutoffs with 
vanishing $H(\Lambda)$, we recover the results of Ref.\cite{kharchenko}.)
Although one three-body datum is needed as input,
the EFT has not lost its predictive power. We can still predict (i) the
energy dependence of $S=1/2$ $nd$-scattering and (ii) the binding
energy of the triton. 

\begin{figure}[htb]
\begin{center}
\epsfxsize=10cm
\centerline{\epsffile{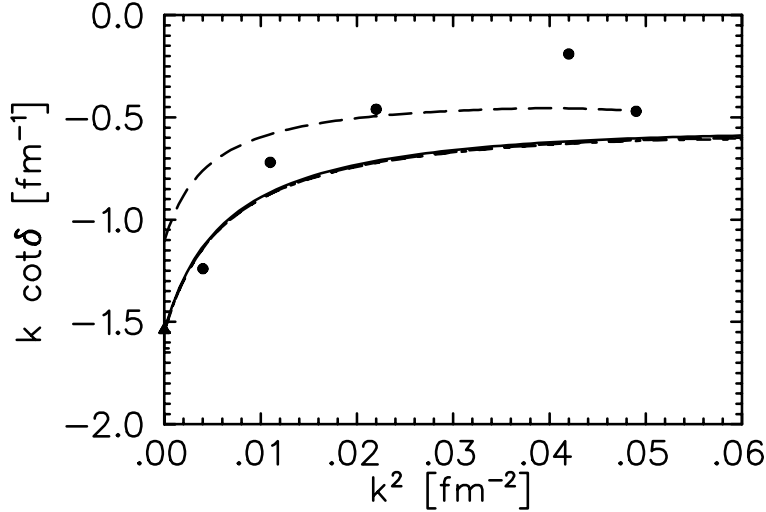}}
\end{center}
\caption{Energy dependence for $S=1/2$ $nd$-scattering with
three different cutoffs $\Lambda=1.9,\, 6.0,\, 11.6 \mbox{ fm}^{-1}$
and $\Lambda_*=0.9\mbox{ fm}^{-1}$ (hidden under the solid curve).
Dashed curve gives estimate of range corrections. Experimental points are 
from van Oers and Seagrave \protect\cite{vOers} (dots) and Dilg et al. 
\protect\cite{dilg} (triangle).}
\label{fig:ed12}
\end{figure}

The resulting energy dependence of $S=1/2$ $nd$-scattering
for three different cutoffs is shown in Fig. \ref{fig:ed12}.
It is clearly seen that the introduction of the three-body force 
renders the low-energy amplitude cutoff independent.
The scattering length is reproduced exactly because it was used
to fix $\Lambda_*$. The agreement for finite momentum is at least
encouraging. Our experience from the $S=3/2$ channel is that
the range corrections improve the agreement considerably (cf. Fig. 
\ref{fig:res32}). The dashed curve in Fig. \ref{fig:ed12} gives a crude 
estimate of these corrections. In the zero range approximation, we 
have the relation
\beq
\label{range}
\sqrt{MB_2}=1/a_2^t, 
\eeq
which holds only approximately  in nature. 
In our calculations we take the deuteron binding energy $B_2$ from experiment 
and determine $a_2^t$ from Eq. (\ref{range}). The dashed curve in Fig.
\ref{fig:ed12} is obtained by taking the experimental value of $a_2^t$
as input and leaving all other parameters unchanged.
From the estimated size of the range corrections, we anticipate an 
improved agreement once the range corrections are included. (One should also
keep in mind that the experimental phase shift analysis \cite{vOers}
does not give any error estimate).

The triton binding energy is obtained
from the solution of the homogeneous equations corresponding to
Eqs. (\ref{aeqf}, \ref{beqf}) for $E=-B_3$.
We find $B_{3} = 8.0$ MeV for the triton binding energy, 
to be compared with the experimental result
$B_3^{exp} = 8.48$ MeV which is known to very high precision. 
For a leading order calculation, the agreement is very good.
The theory without pions seems to work for triton physics. However,
to draw definite conclusions one has to calculate the range corrections
for both the energy dependence and the triton binding energy.

\begin{figure}[htb]
\begin{center}
\epsfxsize=10cm
\centerline{\epsffile{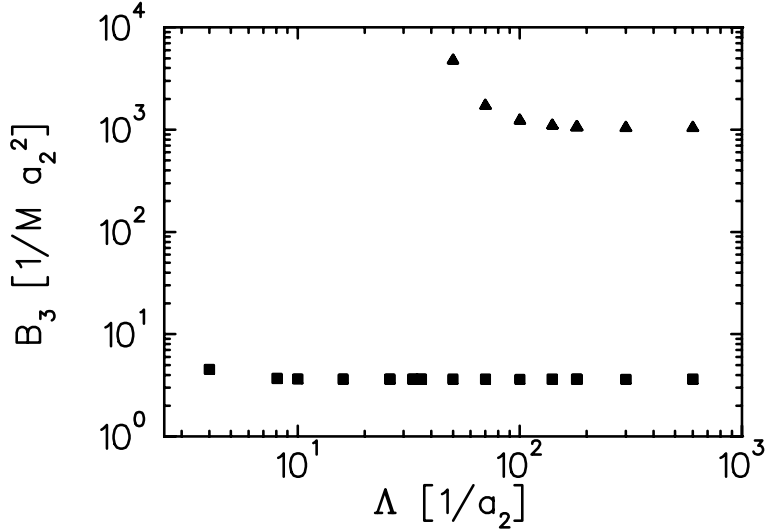}}
\end{center}
\caption{Three-nucleon bound state spectrum for $\Lambda_*=0.9\mbox{ fm}^{-1}$.
The shallowest bound state corresponds to the triton.}
\label{fig:bs12}
\end{figure}

In Fig. \ref{fig:bs12} we show the bound state spectrum as a function of 
the cutoff $\Lambda$. 
The shallowest bound state is the triton. Its binding energy is cutoff
independent. However, as $\Lambda$ is increased
new deeper bound states appear whenever $H(\Lambda)$
goes through a pole. 
\begin{figure}[htb]
\begin{center}
\epsfxsize=10cm
\centerline{\epsffile{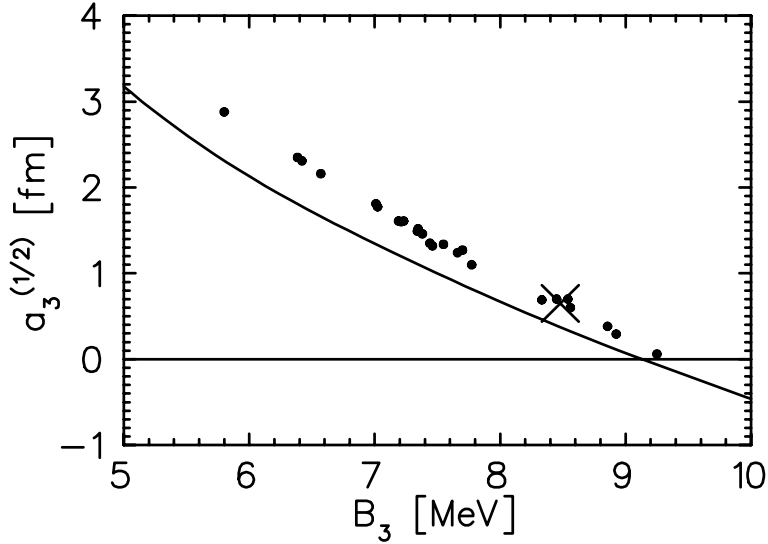}}
\end{center}
\caption{Phillips line: $S=1/2$ $nd$ scattering length $a_3^{(1/2)}$
as function of the triton binding energy $B_3$: EFT to leading order 
(solid line); potential models (dots); experiment (cross).}
\label{fig:phill}
\end{figure}
These new bound states appear with infinite
binding energy directly at the pole. When the cutoff is increased further,
their binding energy reduces and becomes cutoff independent as well. 
The poles of $H(\Lambda)$ can be 
parametrized as
\beq
\label{poles}
\Lambda_n=\underbrace{f(\Lambda_* a_2)}_{{\cal O}(1)}
\exp(n\pi/s_0) a_2^{-1}\,.
\eeq
One counts $n$ bound states for $\Lambda_{n-1} 
< \Lambda < \Lambda_n$. However, only the states between threshold and 
$\Lambda\sim m_\pi \sim 1/r_2$ are within the range of the EFT. By solving 
Eq. (\ref{poles}) for the number of bound states $n$, we then obtain
\beq
\label{efimovspec}
\#_{BS}=\frac{s_0}{\pi}\ln\left(\frac{a_2}{r_2}\right)
+ {\cal O}(1)\,,
\eeq
and recover the well-known Efimov effect.\cite{efimov}
In the limit $a_2 \to \infty$, an infinite number of 
shallow three-body bound states accumulates at threshold.
That these bound states are shallow follows from the fact that 
the binding energy is naturally given in units of 
$1/(M a_2^2)$ which vanishes as $a_2 \to \infty$.
Furthermore, we also recover the Thomas effect.\cite{thomas}
In a hypothetical world where
$\Lambda \sim m_\pi \sim 1/r_2 \to \infty$, the range of the EFT 
increases and deeper and deeper physical bound states appear. 
Consequently, there is an infinitely deep bound
state for $\Lambda \to \infty$. However, for $m_\pi \sim 1/r_2$ as
in the real world the deep bound states are outside the range of the EFT
and their presence does not influence the physics of the shallow ones. 

Moreover, the variation of the parameter $\Lambda_*$ gives a natural 
explanation for the Phillips line.\cite{phillips}
No matter how good the agreement with the actual experimental number is,
values of $B_3$ and $a_3^{(1/2)}$ in different potential models
are correlated and fall on the Phillips line. 
Obviously, there is a correlation
between $B_3$  and $a_3^{(1/2)}$. In the EFT framework,
different models correspond in leading order 
to different values of $\Lambda_*$.
Varying $\Lambda_*$ then generates the observed Phillips line.
The dynamics of QCD chooses a particular value which up to higher order
corrections is $\Lambda_*=0.9\mbox{ fm}^{-1}$.
In Fig. \ref{fig:phill} we show the Phillips line obtained in the 
EFT compared with results from various potential model calculations
\cite{jim} and the experimental values for $B_3$  and $a_3^{(1/2)}$.
Our Phillips line is slightly below the one from the 
potential models. We expect this discrepancy to be reduced
once range corrections are taken into account.

\section{Conclusions}
We have studied the three-nucleon system using EFT methods. While for
$nd$-scattering in the $S=3/2$ channel precise predictions are obtained 
in a straightforward way, the $S=1/2$ channel is more complicated.
It displays a strong cutoff dependence even though all individual
diagrams are UV finite.
In this channel a nonperturbative renormalization takes place 
similar to the case of spinless bosons. This renormalization 
requires an $SU(4)$ symmetric three-body force which is
enhanced to leading order by the renormalization group flow.
Together with the recent results of Mehen et al.\cite{caltech}
this gives an explanation for the approximate $SU(4)$ symmetry in nuclei.
Furthermore, we find that the Phillips line is a consequence of 
variations in the new dimensionful parameter $\Lambda_*$ which is 
introduced by the three-body force. $\Lambda_*$ is not given
from two-nucleon data alone and has to be determined from a 
three-body datum. In the appropriate limits for the two-body 
parameters $a_2$ and $r_2$, 
we also recover the well known Thomas and Efimov effects.

The leading order of the EFT gives a quantitative description of
the triton binding energy and the energy dependence for $S=1/2$
$nd$-scattering. The theory shows the potential for a realistic 
description of the triton once the range corrections are included.
Immediate applications of the EFT include polarization observables 
in $nd$-scattering and triton properties such as its charge form
factor. The incorporation of the long range Coulomb force would widen the 
possible applications considerably as $pd$-scattering and the
physics of $^3$He becomes accessible.

Finally, since the pionless theory is equivalent to the leading order
in KSW counting,\cite{3musketeers}
the success in the three-nucleon system opens the possibility
of applying the EFT method to a large class of systems with three
or more nucleons.

\section*{Acknowledgements}
This work was done in collaboration with P.F. Bedaque and U. van Kolck,
whom I thank for many valuable discussions. 
%and some comments on the manuscript.
I would also like to thank Jim Friar for the potential model data.
This research was supported by the Natural Sciences and Engineering
Research Council of Canada.

\end{document}